\documentstyle[11pt,COIMBRA_pasp,epsf]{article}      

\begin{document}

\title{Atmospheric parameters for non-interacting eclipsing binaries: 
preliminary results from synthetic photometry}

\author{E. Lastennet, F. Cuisinier}
\affil{Depto. de Astronomia, UFRJ, Rio de Janeiro, Brazil}

\author{T. Lejeune}
\affil{Observat\'orio Astron\'omico, Coimbra, Portugal}

\begin{abstract}
Double-lined spectroscopic binaries which are also eclipsing provide the
most accurate stellar data, and are consequently of first importance
to improve stellar evolution modelling.
While the mass and radius of each component of detached eclipsing
binaries (EBs) can be accurately known, their temperature (T$_{\rm eff}$) and
chemical composition ([Fe/H]) are still uncertain.
To contribute to overcome these difficulties, we present
metallicity-dependent T$_{\rm eff}$
determinations of 11 EBs (22 individual
stars) from  Str\"omgren synthetic photometry.
Inconsistencies suggesting new photometric observations are discussed.
Moreover, by exploring the best $\chi^2$-fits to the photometric data, we have
re-derived their reddening.
\end{abstract}


\section{Introduction}
A small sample of detached EBs provide accurate individual
mass and radius (Andersen 1991).
As stressed by many authors (see e.g. Clausen 1993), [Fe/H] is the main
remaining uncertainty of these stars, and their T$_{\rm eff}$
are inhomogeneously determined (Lastennet et al. 1999, LLWB99).
The determination of these two last parameters is motivating
because the knowledge of all possible stellar parameters
for such stars is the basis of the modelling of the global physical
properties and evolution of star clusters or galaxies.
In this context, the present work follows LLWB99
to determine metallicity-dependent T$_{\rm eff}$s of 
non-interacting EBs from Str\"omgren photometry.

\section{Sample of eclipsing binaries and method}
Table 1 gives the individual Str\"omgren photometry and surface gravity we
used for our working sample.
This sample covers a wide mass range, from 1.198 (UX Men B)
to 27.27 M$_{\odot}$ (V3903 Sgr A).
To be consistent with the LLWB99 work
we only kept EBs with  
(b$-$y), 
m$_1$ 
and c$_1$ 
data.
The other commonly used Str\"omgren index $\beta$ -
not affected by interstellar reddening or the distance of the
stars - would provide another observational constraint,
however we decided to exclude it from our study.
The reason is twofold:
first, determinations for both components of a system are not always
available (mainly due to the narrow bandpass of the H$_{\beta}$
filters), and secondly, the synthetic $\beta$ index from the BaSeL
models is not reliable because - by construction - the resolution
of flux distributions is 10-20 $\AA$ (see Lastennet 1998 for a
proposed correction).

\subsubsection{Photometric errors}
While only binaries with photometric errors were considered in LLWB99,
here we consider systems with no photometric errors published,
except for HS Hya
and V906 Sco.
For the case were these errors are
supplied\footnote{
i.e. HS Hya, V906 Sco and
the binaries listed in LLWB99,
excluding IQ Per and YZ Cas because their B components are much fainter
than the A components, implying large errors in their colour indices
measurements},
the mean values are
$\sigma$$_{(b-y)}$$=$0.013, $\sigma$$_{(m1)}$$=$0.022 and
$\sigma$$_{(c1)}$$=$0.027.
We adopt these mean errors for the systems reported with no
photometric errors in Tab. 1.
\scriptsize
\begin{table*}[hbt]
\caption[]{
Basic data
from Jordi et al. 1997 (except HS Hya and V760 Sco)
for the sample.
Reddening is given in the two last columns.
}
\label{tab:data}
\begin{flushleft}
\begin{tabular}{lccccll}
\tableline\noalign{\smallskip}
System & (b$-$y)   &   m$_1$   &  c$_1$ &  $\log$ g  & E(b$-$y)$^\dag$
&  E(b$-$y) \\
\noalign{\smallskip}
\tableline\noalign{\smallskip}
V539 Ara     & $-$.038 &   .088 &   .249  &  3.926$\pm$0.017 &  0.050 & 0.051$^{(1)}$ \\
             & $-$.032 &   .090 &   .285  &  4.096$\pm$0.022 &  0.050 & 0.053$^{(1)}$ \\
QX Car       & $-$.072 &   .087 &   .036  &  4.140$\pm$0.020 &  0.032 & 0.036$^{(2)}$ \\
             & $-$.072 &   .092 &   .076  &  4.151$\pm$0.021 &  0.029 &              \\
SZ Cen       & .188 &   .210 &   .983  &  3.486$\pm$0.008 &  0.000 & 0.058$^{(3)}$ \\
             & .166 &   .188 &  1.019  &  3.677$\pm$0.007 &  0.051 &                \\
$\chi^2$ Hya & $-$.02  &   .11  &   .83   &  3.712$\pm$0.015 &  0.013 & 0.012$^{(4)}$ \\
             & $-$.01  &   .11  &   .84   &  4.188$\pm$0.019 &  0.025 &               \\
UX Men       & .359 &   .161 &   .371  & 4.272$\pm$0.009 &  0.070 & 0.02$\pm$0.02$^{(5)}$  \\
             & .368 &   .174 &   .367  & 4.306$\pm$0.009 &  0.070 &              \\
V760 Sco     & .155 &   .029 &   .373  & 4.177$\pm$0.021 & 0.230 & 0.24$^{(6)}$  \\
             & .162 &   .027 &   .410  & 4.259$\pm$0.019 &  0.240 &               \\
V1647 Sgr    & .022 &   .163 &  1.018  & 4.253$\pm$0.012 &  0.029 & 0.029$^{(7)}$ \\
             & .057 &   .182 &  0.979  & 4.289$\pm$0.012 &  0.030 & 0.030$^{(7)}$ \\
V3903 Sgr    & .184 &   .006 & $-$.114 & 4.058$\pm$0.016 &  0.310 & 0.32$^{(8)}$  \\
             & .191 &   .001 & $-$.076 & 4.143$\pm$0.013 &  0.310 &               \\
CV Vel       &$-$.067 & .100 &   .269  &  4.000$\pm$0.008 &  0.013 & 0.030$^{(9)}$ \\
             &$-$.064 & .097 &   .277  &  4.023$\pm$0.008 &  0.018 &  \\
HS Hya       &.289$\pm$.007  & .144$\pm$.007 & .421$\pm$.007 & 4.3259$\pm$0.0056 & 0.000 &  $-$0.004$^{(10)}$ \\
             &.302$\pm$.007  & .156$\pm$.008 & .374$\pm$.007 & 4.3539$\pm$0.0057 & 0.000 &                  \\
V906 Sco     &.044$\pm$.003 & .126$\pm$.004 & 1.023$\pm$.005 & 3.656$\pm$0.012 & 0.070 & 0.059$^{(11)}$  \\
             &.063$\pm$.002 & .094$\pm$.002 & 1.183$\pm$.002 & 3.858$\pm$0.013 & 0.093 &                 \\
\noalign{\smallskip} \tableline
\tableline
\noalign{\smallskip}
\noalign{\smallskip}
\end{tabular}
\\
\scriptsize
 $^\dag$ this  work.
$^{(1)}$ Clausen (1996, \aap, 308, 151);
$^{(2)}$ Andersen et al. (1983, \aap, 121, 271);
$^{(3)}$ Gr{\o}nbech et al. (1977, \aap, 55, 401);
$^{(4)}$ Clausen \& Nordstr\"{o}m (1978, \aap, 67, 15);
$^{(5)}$ Andersen et al. (1989, \aap, 211, 346);
$^{(6)}$ Andersen et al. (1985, \aap, 151, 329);
$^{(7)}$ Andersen \& Gim\'enez (1985, \aap, 145, 206);
$^{(8)}$ Vaz et al. (1997, \aap, 327, 1094);
$^{(9)}$ Clausen \& Gr{\o}nbech (1977, \aap, 58, 131);
$^{(10)}$ Torres et al. (1997);
$^{(11)}$ Alencar et al. (1997, \aap, 326, 709).
\end{flushleft}
\end{table*}
\normalsize
\vspace{-0.4cm}
\subsubsection{Method}
We apply a $\chi^2$-minimization method on the BaSeL models (Lejeune et al.
1998, see also Lastennet, Lejeune \& Cuisinier, these proceedings)
to derive the T$_{\rm eff}$ and [Fe/H]
values matching simultaneously the observed Str\"omgren
photometry,
the surface gravity (log g) being fixed to its
accurately determined value (see LLWB99
for details).

\section{Discussion}
\subsubsection{Results derived from (b$-$y), m$_1$ and c$_1$}
The surprising result is that
BaSeL is unable
to match simultaneously b$-$y, m$_1$ and c$_1$,
independently of the value of reddening adopted.
\begin{figure}
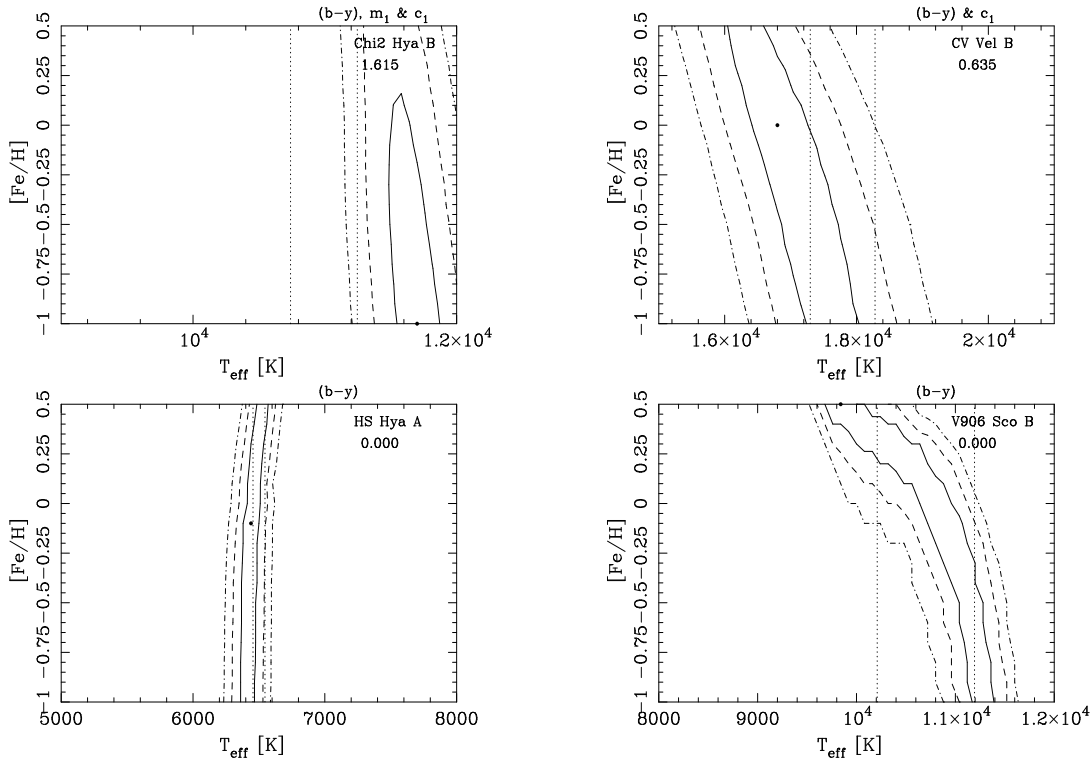

\centerline{
\plotfiddle{coimbra2_fig1.ps}{5.cm}{-90}{24}{25}{-220}{140}
\plotfiddle{coimbra2_fig2.ps}{5.cm}{-90}{24}{25}{-380}{140}
}
\centerline{
\plotfiddle{coimbra2_fig3.ps}{5.cm}{-90}{24}{25}{-220}{140}
\plotfiddle{coimbra2_fig4.ps}{5.cm}{-90}{24}{25}{-380}{140}
}
 \caption{
 \small
Examples of (T$_{\rm eff}$-[Fe/H]) solutions matching b$-$y, m$_1$ and c$_1$
 (upper left) and b$-$y and c$_1$
 (upper right).
T$_{\rm eff}$ determination from the b$-$y index alone:
HS Hya A and V906 Sco B (lower panels).
 Previous determinations are shown as vertical lines (references
 in Ribas et al. 2000).
 \normalsize
 }
\end{figure}
All the $\chi^2$-scores are bad ($>$ 10), except for both components
of the system $\chi^2$ Hya: its primary component gives a very good agreement
with previous T$_{\rm eff}$ determinations, but we
favour a T$_{\rm eff}$ larger by $\sim$ 700K for $\chi^2$ Hya B (see Fig. 1).
Of course, the $\chi^2$-scores are related to the assumed photometric errors,
so largest errors should give better (i.e. smaller) $\chi^2$ values.
The 2 binaries with published photometric errors (HS
Hya and  V760 Sco) give bad fit as well.
Further study is needed, as well as more information on the observed
Str\"omgren data before definitive conclusion.
Reasons for these discrepancies could be that:
{\bf 1)}
the BaSeL models are not fit to predict Str\"omgren indices.
This explanation is possible because some problems
due to the m$_1$ and c$_1$ indices were detected for F-type stars with the
BaSeL Str\"omgren synthetic photometry (Lastennet et al.
2001).
However, only UX Men and HS Hya contain F-type stars, so this cannot
be a general explanation for our sample.
Moreover, only $\sim$17\% (7 over 40) of the stars studied by LLWB99
presented  a similar bad fit (and this was not correlated with a
particular T$_{\rm eff}$ range) against $\sim$91\% (20 over 22) in the
present sample.
Another convincing point against unreliable results from the BaSeL
Str\"omgren synthetic photometry is that
among the 9 F-type stars studied by LLWB99,
only 2 stars show bad fits.
{\bf 2)}
the differential reddening in the direction of these stars is
strongly different from the standard values that we adopt.
A comparison between the reddening found in the literature
and derived from BaSeL
(Tab. 1) show a good agreement,
except for SZ Cen A but even in this case the disagreement
is only of 0.058 mag.
{\bf 3) }
the choice of b$-$y, m$_1$ and c$_1$ is not critical enough
for the purpose of this work. When it is true that b$-$y becomes
increasingly insensitive to T$_{\rm eff}$ for the hotter stars
(for (b$-$y)$_0$$<$0), it appears that even in this defavorable range
BaSeL is able to predict good results (e.g. EM Car or CW
Cep in LLWB99).
It is worth noting that even bad fits give results
in agreement with previous studies.
Another unexpected (because badly fitted)
but interesting result is obtained for QX Car: we predict a rich metallicity
([Fe/H]$>$0.30 (0.15) from the primary (secondary))
which is in agreement with previous estimates:
Lastennet (1998) (Z$=$0.04 from 2 different
stellar models, i.e. [Fe/H]$\sim$0.39), and the extrapolated result of Ribas
et  al. (2000) (Z$=$0.035, i.e. [Fe/H]$\sim$0.32).
{\bf 4)}
the observed colors of these stars are in some way erroneous and
should be carefully re-determined from new photometric observations.

\subsubsection{T$_{\rm eff}$ and/or [Fe/H] from 1 or 2 photometric
constraints}
When bad fits were obtained using all the photometric data (3 colours),
LLWB99
considered the solutions derived from the
combination of 2 colours.
In this case, the results match - all with
success -  (b$-$y) and c$_1$ simultaneously (e.g. Fig. 1, right upper panel).
A comparison of T$_{\rm eff}$(BaSeL) with previous
studies shows a general good agreement, but T$_{\rm eff}$(BaSeL)
are slightly but systematically lower.
Unfortunately, few information is derived for [Fe/H], all the range
considered being virtually possible inside the 1-$\sigma$ contours.
Finally, we show in Fig. 1 (lower panels) the solutions obtained from
the b$-$y index alone for the 2 only systems with photometric errors.
The T$_{\rm eff}$
determinations show a perfect agreement with previous works.
A precise spectroscopic determination of the HS Hya metallicity is
needed  but if one assumes the value of [Fe/H]$=-$0.17 (Torres et al., 1997),
then  we predict T$_{\rm eff}$ in the range 6380-6440 K (primary) and
6320-6380 K  (secondary). These T$_{\rm eff}$s are slightly lower than the
results  of Torres et al.: 6450-6550 (primary) and 6350-6450 K (secondary).

\end{document}